%% file: main.tex
\documentclass[12pt]{article}
\DeclareUnicodeCharacter{2212}{-}
\usepackage[utf8]{inputenc}
\usepackage{fullpage}
\usepackage{hyperref}
\usepackage{graphicx}
\usepackage{lipsum}
\usepackage[sort&compress,numbers]{natbib}
\usepackage{hyperref}
\hypersetup{hidelinks}
\usepackage[total={6.5in,9in},top=1in,headsep=0.1in,headheight=1in]{geometry}
\usepackage{siunitx}
\usepackage{enumitem}
\setenumerate{itemsep=0mm}

\newcommand\snowmass{
\begin{center}
  \rule[-0.2in]{\hsize}{0.01in}\\
  \rule{\hsize}{0.01in}\\
  \vskip 0.1in
  Submitted to the Proceedings of the US Community Study\\ 
  on the Future of Particle Physics (Snowmass 2021)\\
  \rule{\hsize}{0.01in}\\
  \rule[+0.2in]{\hsize}{0.01in}\\[-2em]
\end{center}
}

\usepackage[firstpage=true]{background}
\backgroundsetup{contents={\parbox{6.5in}{\snowmass}}, scale=1,placement=top,opacity=1,color=black,position={3.25in,1.2in}}

\usepackage{fancyhdr}
\fancypagestyle{plain}{%
  \fancyhf{}%
  \fancyhead[C]{}
  \fancyfoot[C]{\thepage}
}

\fancypagestyle{empty}{%
  \fancyhf{}%
  \fancyhead[C]{{\it Primordial Black Hole Dark Matter}}
  \fancyfoot[C]{\thepage}
}
\title{Snowmass2021 Cosmic Frontier White Paper:\\
Primordial Black Hole Dark Matter}
\date{}

\usepackage{authblk}

\usepackage{aas_macros}
\input{commands.tex}
\input{authors.tex}

\definecolor{offblue}{rgb}{0.26,0.4,0.74}
\definecolor{dkred}{rgb}{0.5,0,0}

\begin{document}

\maketitle

\begin{abstract}


Primordial Black Holes (PBHs) are a viable candidate to comprise some or all of the dark matter and provide a unique window into the high-energy physics of the early universe.
This white paper discusses the scientific motivation, current status, and future reach of observational searches for PBHs.
Future observational facilities supported by DOE, NSF, and NASA will provide unprecedented sensitivity to PBHs.
However, devoted analysis pipelines and theoretical modeling are required to fully leverage these novel data.
The search for PBHs constitutes a low-cost, high-reward science case with significant impact on the high energy physics community.

\end{abstract}

\section{Executive Summary}

The nature of dark matter and the physics of the early universe are two high priority science cases within the Cosmic Frontier of High Energy Physics.
As potentially the first density perturbations to collapse during the early universe, primordial black holes (PBHs) present our earliest window into the birth of the universe and energies between the QCD phase transition and the Planck scale.
The corresponding length scales ($k = 10^{7} - 10^{19}$ $h\,\mathrm{Mpc}^{-1}$) are much smaller than those measured by other current and future cosmological probes.
While earlier estimates suggested that much of the PBH dark matter parameter space was constrained, more sophisticated analyses have relaxed many of these constraints, opening up the possibility that PBHs in certain mass ranges comprise the entirety of dark matter.


The detection of PBHs would immediately change our understanding of the physics of the early universe.
This significant reward motivates the further development of several complementary techniques that are sensitive to PBHs and subject to fewer astrophysical systematics, such as gravitational microlensing, gravitational wave detection, and gamma-ray signatures of PBH evaporation. 
Fortunately in many cases PBH science can be done by already well-justified (and often under construction) facilities, e.g.~LIGO, LISA, Rubin, Roman, SKA, MeV gamma-ray facilities, and imaging air Cherenkov telescopes. 
That said, realising PBH science from these facilities requires both dedicated data analysis and theoretical studies.
In addition, it is important to more closely integrate DOE HEP science efforts with enabling NSF and NASA facilities.\\



\noindent {\bf Key Opportunities:}
\begin{enumerate}

    \item Current and near-future observations can provide unprecedented sensitivity to PBH physics. However, it is necessary to ensure that these facilities acquire their data with a cadence and sky coverage that enables PBH searches \citep[e.g.~][]{drlica-wagner_2019_lsst_dark_matter,1812.03137,street_2018b}.

    \item The sensitivity of PBH searches will be maximized by combining data sets from multiple observational facilities. Development of joint processing and analyses of Rubin Observatory, Roman Space Telescope, and Euclid will maximize the opportunity to detect PBHs. 
    
    \item Current and future gravitational wave facilities will provide an unparalleled opportunity to detect PBHs directly through gravity. These facilities should be supported by the HEP community and include both ground-based detectors such as LIGO and Cosmic Explorer and space-based such as LISA and AEDGE.
    
    \item The scale of current and near-future data sets and the complexity of PBH analyses will benefit from collaborative scientific teams. These teams will develop the tools to perform rigorous and sensitive searches for PBHs in current and near-future observational data. The computational challenges presented by these searches are well-matched to the capabilities of HEP scientists and facilities.

    \item Theoretical research will help us better understand the production mechanisms, clustering, and spin properties of PBH. These characteristics will inform the expected abundance of black hole microlensing/GW events and systematics with cosmic surveys, as well as PBH connections to primordial physics.
    Furthermore, improved simulations of the PBH merger rate and of PBH-specific accretion rates will help inform observational constraints.

\end{enumerate}

\section{Introduction}

\begin{figure}
\begin{center}
\includegraphics[width=0.9\hsize]{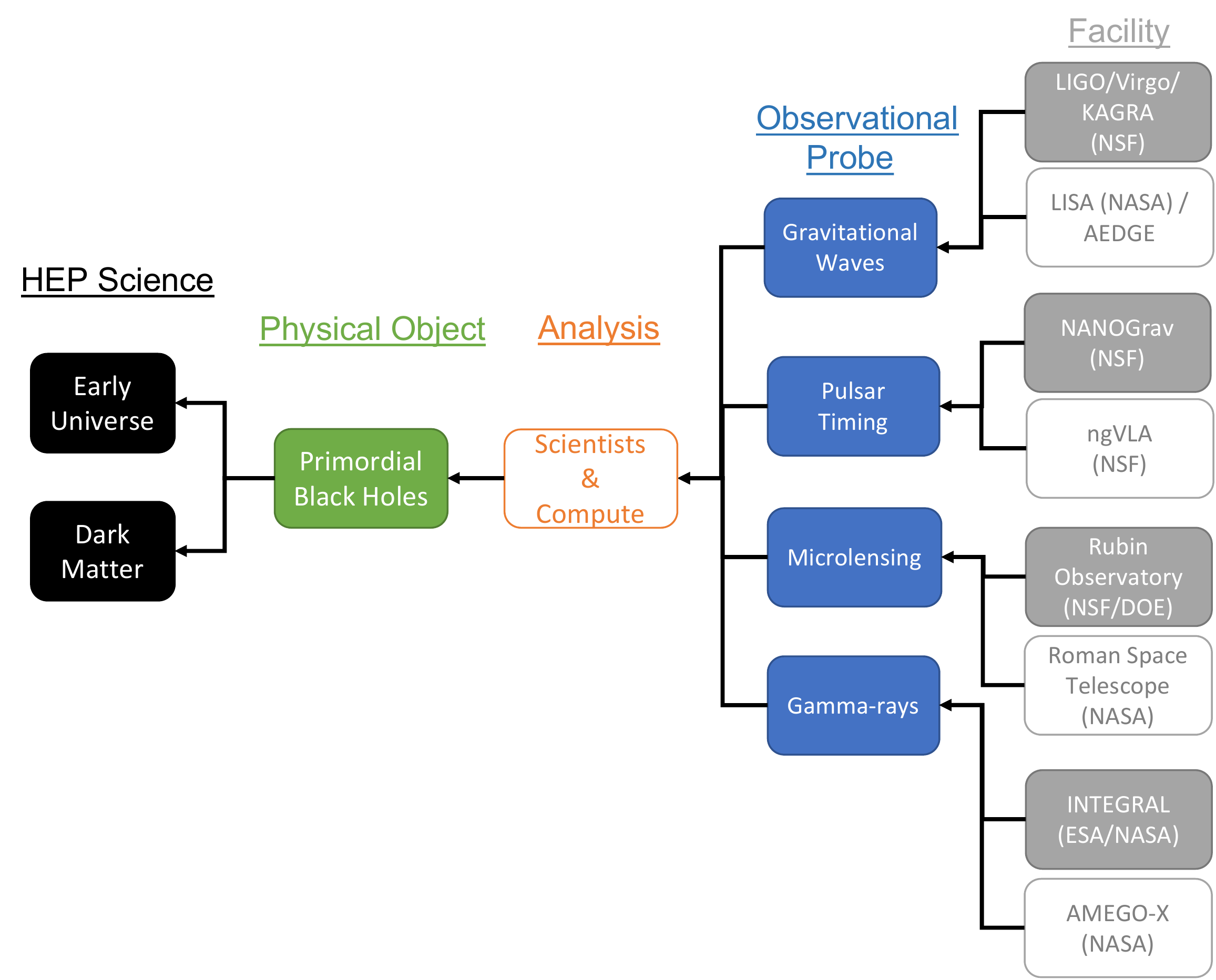}
\end{center}
\caption{
Conceptual map relating DOE/HEP early universe and dark matter science interests to various observational probes and facilities, via primordial black holes. Shown are the general observational techiques as well as specific facilities relying on them. Current facilities are shown by the filled grey boxes, while planned facilities are represented by the non-filled boxes. Specific recommendations are for scientist support and compute necessary to turn the observational data into scientific results. For facilities we denote current U.S. sponsors that DOE could partner with in parentheses.
}
\label{fig:overview}
\end{figure}

\subsection{Primordial Black Holes as Dark Matter}

\begin{figure}
\begin{center}
\includegraphics[width=0.9\hsize]{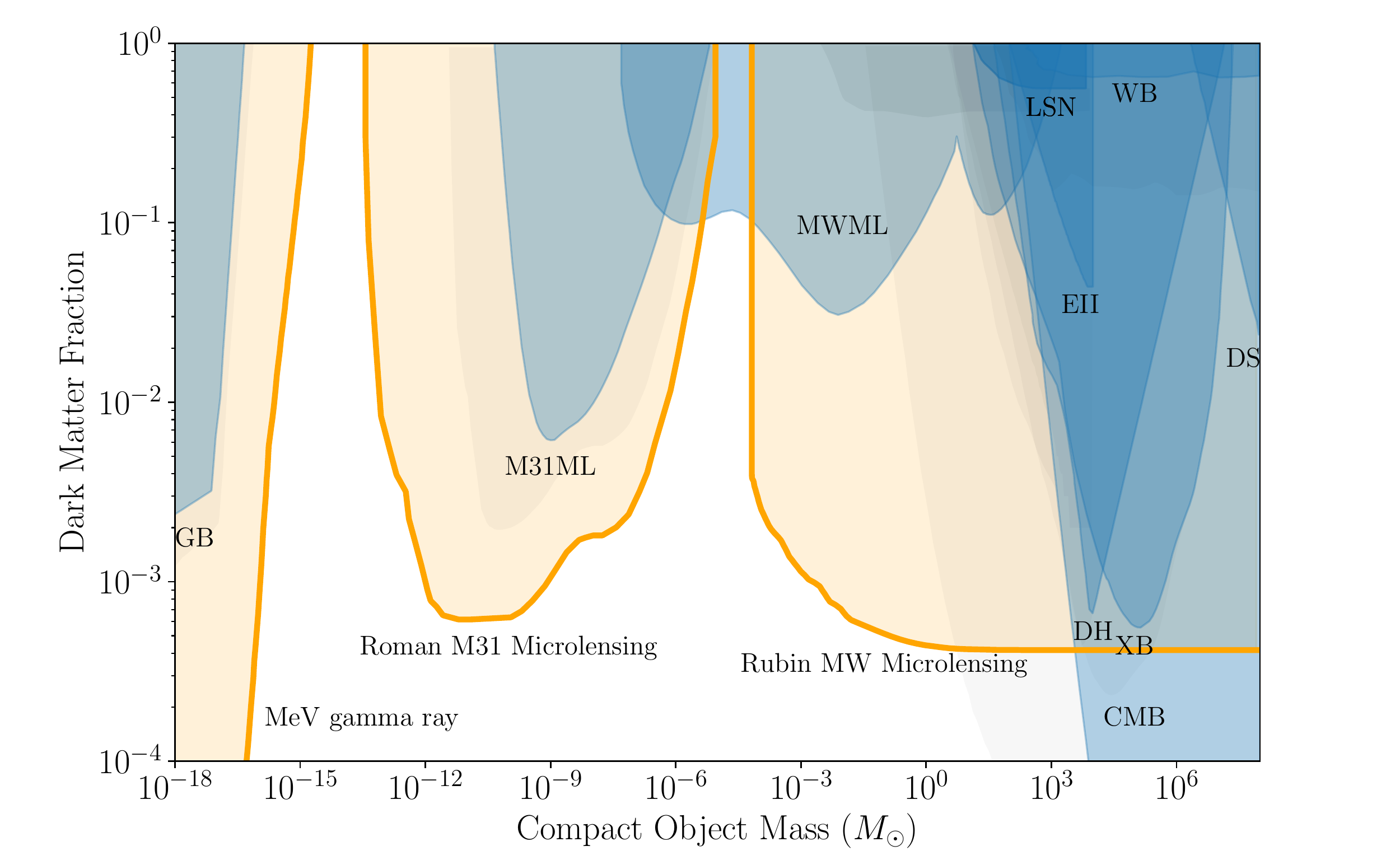}
\end{center}
\caption{
Current constraints on dark matter fraction composed of compact objects of a given mass (blue and gray) and selected projections for future gamma-ray and microlensing probes (gold). The blue and gray regions denote constraints based on more and less conservative assumptions, respectively. Forecast projections for Rubin or Roman dedicated microlensing surveys of M31 from Ref.~\citep{drlica-wagner_2019_lsst_dark_matter} and for MeV gamma ray facilities from Ref.~\cite{Coogan:2020tuf} are displayed.
The existing constraints are from M31 microlensing (M31ML)~\cite{Niikura:2019kqi}, MACHO/EROS microlensing (MWML) \cite{MACHO_2001, Tisserand_2007, Wyrzykowski_2011, 2022arXiv220213819B}, supernovae lensing (LSN)~\cite{Zumalacarregui:2017qqd}, Eridanus II dwarf galaxy (EII)~\cite{Brandt:2016aco,DES:2016vji}, wide binary stars (WB)~\cite{Quinn:2009zg,Yoo:2003fr}, dwarf galaxy dynamical heating (DH)~\cite{Lu:2020bmd,Takhistov:2021aqx,Takhistov:2021upb}, x-ray binaries (XB)~\cite{Inoue:2017csr}, CMB distortions from accreting plasma by PBHs in early universe (CMB)~\cite{Ali-Haimoud:2016mbv,Ricotti:2007au}, and disk stability (DS) constraints~\cite{Xu:1994vb}.
Note that recent work by \cite{golovich:2020} suggests that it is challenging to extend the Rubin microlensing probes beyond $10^3 M_\odot$.
}
\label{fig:dm}
\end{figure}

Compact objects, particularly black holes, are one of the oldest models of dark matter. PBHs are the modern realisation of MAssive Compact Halo Objects (MACHOs),
constraints on which provided the first \emph{direct} constraints on the nature of dark matter in the 1990s \citep{1992ASPC...34..193A,1992AcA....42..253U, 1993Msngr..72...20A}. Current microlensing constraints set limits on the black hole abundance at the level of $10\%$ for black holes in the mass range $0.01 - 100 \Msun$ (see Figure~\ref{fig:dm}) \cite{MACHO_2001,Tisserand_2007,Wyrzykowski_2011, 2022arXiv220213819B}. The science cases detailed in this white paper thus have a rich heritage and are backed by a substantial body of literature.

Primordial black holes (PBHs) could form at early times from the direct gravitational collapse of large density perturbations that originated during inflation.
The same fluctuations that lay down the seeds of galaxies, if boosted on small scales, can lead to some small areas having a Schwarzschild mass within the horizon, which spontaneously collapse to form black holes \citep{Carr:1974nx,Meszaros:1974,1975Natur.253..251C,Bellido:1996,2016PhRvD..94h3504C} (as recently confirmed with numerical relativity \cite{deJong:2021bbo}). Non-detections of PBH are thus also constraints on the small-scale power spectrum, as shown in Figure~\ref{fig:eu}. However, collapse of horizon-scale perturbations is not the only mechanism for PBH formation. As detailed in Section~\ref{sec:subhorizon}, they may also be formed through collapse and fragmentation of high energy scalar fields via isocurvature fluctuations, and may be generic in super-symmetry.  


Compact object dark matter is fundamentally different from particle models; primordial black holes 
cannot be created in an accelerator and can only be detected observationally. Much of the parameter space has been constrained by existing probes, but a large window remains where PBHs around asteroid mass ($10^{-15}$ to $10^{-10} M_\odot$) could make up the entirety of dark matter.
Primordial black holes are also a plausible source of the population of merging black holes around $30 \Msun$ recently detected by LIGO \citep{Bird:2016, Clesse:2016, Sasaki:2016}, a mass range where PBHs may still comprise $\gtrsim \mathcal{O}(10)\%$ of the dark matter.

Limits on the abundance and mass range of primordial black holes are necessarily observational. At order of magnitude, the black hole mass is set by the mass enclosed within the horizon at the time of black hole collapse and thus ranges between $10^{-18} \Msun$ ($10^{15}$ g), below which the black hole would evaporate, and $10^9 \Msun$ ($10^{42}$ g), above which structure formation, Big Bang Nucleosynthesis and the formation of the microwave background would be severely affected \citep{Sasaki:2018}. 

Observational limits on primordial black holes have evolved over time: numerous constraints have been proposed based on detailed modelling and then revised as the modelling has been updated. For example, early CMB constraints on PBHs \citep{Ricotti:2008} were found to rely on an overly optimistic model for the black hole accretion rate \citep{AliHaimoud:2017cmb} and continue to be updated as our understanding of accretion improves \citep{Serpico:2002.10771}. Similarly, uncertainty in the modelling of the PBH merger rate has led to several revisions of the bounds from gravitational wave events (see Section~\ref{sec:gwmerger}). Finally, initially strong dynamical constraints from dwarf halos \citep{2016ApJ...824L..31B, 2017PhRvL.119d1102K} were revised and weakened by higher fidelity Fokker-Planck modelling \citep{Zhu:2017plg} and then strengthened by a combined statistical analysis \citep{Stegmann:2019wyz}. Current dynamical limits appear to rule out PBHs being all the dark matter, but strong modelling assumptions remain \citep{Zhu:2017plg, Clesse:2017, Stegmann:2019wyz}.

For stellar mass black holes, the gold standard for detecting compact objects is microlensing. As we shall discuss later, the Vera Rubin Legacy Survey of Space and Time (LSST) will revolutionize microlensing, constraining the abundance of PBHs to $\sim 10^{-4}$ of the dark matter over a wide range of masses \cite{drlica-wagner_2019_lsst_dark_matter,Winch:2020}. Microlensing opportunities in the next ten years are not limited to LSST, and are in particular also provided by the Nancy Grace Roman Space Telescope and the ground-based Extremely Large Telescopes.

\subsection{Primordial Black Holes as Probes of the Early Universe} 
\label{sec:formation}

PBHs are unique amongst black holes in that they form at a high energy density in the early universe, long before  stars, galaxies or the microwave background. Constraints on their abundance are thus also constraints on possible physics at high energies, such as inflation. There are several possible scenarios for PBH formation that lead to different masses, spins, and clustering properties of primordial black holes. These production scenarios can be grouped into two classes: those that produce PBHs with masses close to the horizon mass at the time of formation, and those that produce PBHs with subhorizon masses.

\subsubsection{Formation of horizon-mass PBHs}
\label{sec:horizon}
If inflationary dynamics were to enhance primordial density fluctuations on scales much smaller than those probed by the CMB, entire horizon-sized regions can collapse into black holes once the perturbation re-enters the horizon. 
The PBH mass in these scenarios is proportional to the horizon size at formation; thus lower mass PBHs probe higher energies. 
Since these PBHs form directly from the primordial density fluctuations, a measurement of their abundance would directly constrain the amplitude of density fluctuations \citep{Carr:1974nx, Meszaros:1974}. 
These constraints probe small scales between $k = 10^{7} - 10^{19}$ $h$/Mpc, much smaller than those measured by other current and future probes \citep{Bringmann:2012}. Because these scales are highly non-linear in the late-time universe, there is no other possible constraint; the information present at early times has been washed away by gravitational evolution. 

\begin{figure}
\begin{center}
\includegraphics[width=0.9\hsize]{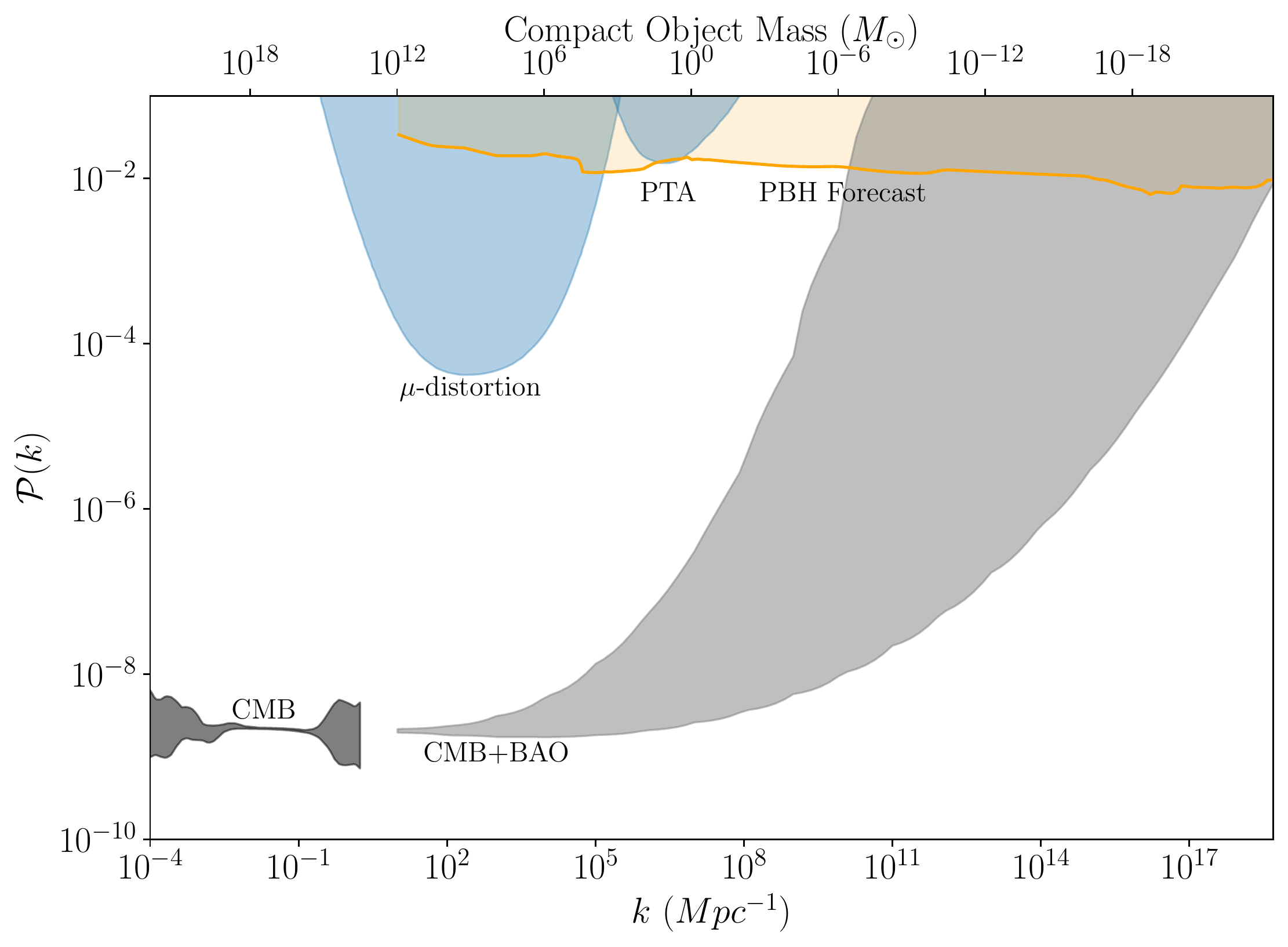}
\end{center}
\caption{
Evidence for or against primordial black holes enables the constraint of the primordial power spectrum to much higher $k$ (i.e., earlier times) than any other probe. 
Observations from the CMB power spectrum bound the primordial power spectrum to be $\mathcal{P}(k) \sim 10^{-9}$ at $10^{-4}\lesssim k \lesssim 10^1\, \mathrm{Mpc}^{-1}$ (black region).
However, at smaller scales, current PBH upper limits from the Pulsar Timing Array (PTA), CMB $\mu$-distortions, and searches for PBHs set significantly weaker constraints.
Based on figures from \cite{Gow_2021, Bringmann:2012} and references therein.
Light grey band shows the power spectrum expected by extrapolating the best-fit CMB power spectrum using a model with higher derivatives $\mathcal{P}_s(k) = A_s (k/k_*)^{n_s - 1 + \frac{1}{2} \alpha_s \ln(k/k_*) + \frac{1}{6} \beta_s \ln(k/k_*)^2}$. The large range of scales probed by PBHs allow for the best constraints on the highest order derivative \citep{Li:2018iwg}. These higher order derivatives correspond to higher orders of the inflationary slow-roll parameters, and so non-zero values are generic in inflationary models. 
}
\label{fig:eu}
\end{figure}

The typical curvature perturbations from inflation are expected to be nearly scale invariant and are constrained by the CMB at large scales. Formation of PBHs in these classes of models requires dramatically enhancing the perturbations on some smaller scale, by hybrid inflation \cite{Bellido:1996}, introducing new features within the inflaton potential~\cite{Garcia-Bellido:2017mdw, Kawasaki:2018daf,Germani:2018jgr, Ashoorioon:2019xqc, Ashoorioon:2020hln}, or through inhomogeneous baryogenesis \cite{Dolgov:1992pu,Dolgov:2008wu,Hasegawa:2018yuy,Kawasaki:2012kn}. Vacuum bubbles formed during inflation can appear to the outside observer as PBHs~\cite{Garriga:2015fdk,Deng:2017uwc,Kusenko:2020pcg, Maeso:2021xvl}. 
Figure~\ref{fig:eu} shows constraints from PBHs on the power spectrum, compared to those from the CMB, stochastic background constraints from pulsar timing arrays and constraints from $\mu$-distortions. Constraints are taken from \cite{Gow_2021, Bringmann:2012}. PBH constraints are shown in gold, as they still depend on several modelling assumptions. 

Reflecting the need to enhance the power spectrum, current constraints are several orders of magnitude weaker than constant extrapolations from the microwave background to these scales. However, PBH constraints are able to constrain higher derivatives of the power spectrum, and thus higher order inflationary slow-roll parameters. In particular they provide the best current limit on the third derivative of the inflationary power spectrum \citep{Li:2018iwg}. 
As shown in Figure~\ref{fig:eu}, this technically rules out the maximum likelihood Planck power spectrum (although the tension is within the $1-\sigma$ error bars).

Enhanced small-scale density fluctuations may also produce detectable GWs, through second-order perturbations~\cite{Saito:2008jc,Bugaev:2009zh,Saito:2009jt,Bugaev:2010bb,Clesse:2018ogk,Domenech:2021wkk} or non-Gaussianity \cite{PhysRevLett.122.201101}. A power spectrum with $\mathcal P_\zeta \sim \mathcal O(10^{-2})$ on small scales produces induced GWs with an energy density $\Omega_\text{GW} h^2 \simeq \mathcal O(10^{-8})$ at the peak frequency. For PBHs with $\sim 30 M_\odot$, in the  LIGO/Virgo range, the peak frequency is $\sim 10^{-9}$\,Hz, slightly smaller than the frequencies constrained by current pulsar timing array experiments (PPTA, EPTA, NANOGrav)~\cite{Inomata:2016rbd,Orlofsky:2016vbd}. However, PBHs in the currently unconstrained mass range $\sim 10^{-16}-10^{-10}M_\odot$ have a peak frequency around $\sim 10^{-4}-10^{-1}$\,Hz, at an amplitude detectable by space-based gravitational wave detectors such as LISA or AEDGE/DECIGO~\cite{Garcia-Bellido:2017aan,Inomata:2018cht,Bartolo:2018rku}. 
This induced GW signal provides a constraint on the shape of the small-scale power spectrum, and so may in future provide constraints on inflation.

Primordial black holes are thus a probe of primordial density fluctuations in a range that is inaccessible to other techniques~\citep{Josan:2009,Bellido:2017,Bellido:2018}. These curvature fluctuations are imprinted on space-time hypersurfaces during inflation, at extremely high energies, beyond those currently accessible by terrestrial and cosmic accelerators. Our understanding of the universe at these high energies, of order $10^{15} $ GeV and above, comes predominantly from extrapolations of known physics at the electroweak scale. Measurements of the primordial density fluctuations via the abundance of primordial black holes would provide unique insights into physics at these ultra-high energies.

\subsubsection{Formation of subhorizon-mass PBHs}
\label{sec:subhorizon}


The overdensity needed for formation of a black hole in the early universe does not have to come from the same source as the primordial fluctuations that lead to large-scale structure.  Scalar fields (such as the Higgs boson or the scalar fields predicted by supersymmetry) are known to have instabilities caused by scalar self-interaction, which could typically~\cite{Kusenko:2019kcu} be much stronger than the force of gravity on short scales~\cite{Khlopov:1985jw,Kusenko:1997si}.  Such instabilities can cause formation of field lumps that become the building blocks of PBHs~\cite{Cotner:2016cvr,Cotner:2017tir,Cotner:2018vug,Cotner:2019ykd}.  From the point of view of field theory, the scalar lumps could be non-topological solitons known as Q-balls~\cite{Coleman:1985ki}, long-lived metastable objects known as oscillons~\cite{Gleiser:1993pt,Amin:2011hj}, or confined heavy quarks \cite{Dvali:2021byy}.  Once formed, the lumps redshift like matter and can come to dominate the energy density of the Universe, at which point gravitational instability can lead to formation of PBHs~\cite{Cotner:2016cvr,Cotner:2017tir,Cotner:2018vug,Cotner:2019ykd}. 

Another possibility is that the density perturbations grow due to attractive interactions other than gravity, such as the scalar interactions which act as ``long-range'' forces on the length scales smaller than the Compton wavelength of the scalar particle.   Since the scalar interactions may be stronger than gravity at these high energies, the rapid growth of isocurvature perturbations is possible even during the radiation dominated era~\cite{Amendola:2017xhl,Savastano:2019zpr,Flores:2020drq}.  While the small clusters of particles formed this way could virialize and exist indefinitely in the absence of dissipation~\cite{Savastano:2019zpr}, the very same scalar interactions that cause the clusters to form provide a dissipation channel via emission of the scalar waves~\cite{Flores:2020drq}.  A simple system with a heavy fermion and a scalar interacting via the Yukawa coupling can lead to enough PBH formation to account for all dark matter~\cite{Flores:2020drq,Flores:2021jas}. The dark matter abundance in this case is related to the baryon asymmetry of the universe, which offers a potential explanation for the ``coincidence problem'', that is, why the densities of dark matter and ordinary matter are within one order of magnitude of each other~\cite{Flores:2020drq,Flores:2021jas}.

It is intriguing that supersymmetry, which is widely considered a plausible candidate for new physics, provides all the conditions for dark matter in the form of primordial black holes. Supersymmetric generalizations of the standard model predict a large number of scalar ``flat directions'' in the potential. The fields parameterizing the flat directions develop a large expectation value during inflation.  The subsequent evolution of the field generically leads to fragmentation of the scalar condensate into lumps, which can collapse into PBHs. The mass scale of the resulting black holes, $(M_{\rm Planck}^3/\Lambda_{\rm SUSY}^2)\sim 10^{23}\, {\rm g}$, naturally falls in the allowed range for dark matter~\cite{Cotner:2019ykd,Flores:2021jas}.


\subsection{Primordial Black Holes as Supermassive Black Hole Seeds}
\label{sec:smbh}


The number of quasars observed above $z=6$ has exceeded 200, with several discovered at $z>7$ \cite{mortlock11,banados18}, including a $z = 7.5$ quasar with a 1.5$\times 10^{9}$M$_{\odot}$ black hole \cite{yang20}. The mechanism by which the super massive black holes (SMBHs) driving these quasars are assembled within a Gyr of the Big Bang is poorly understood. A standard picture of Eddington-limited accretion of massive Pop III seed black holes cannot reasonably produce the SMBHs seen at high redshift, so super-Eddington growth or an increase in the mass of the black hole seeds are often touted as potential solutions \cite{inayoshi19}.

Invoking primordial black holes (PBHs) as seeds for these high redshift SMBHs can plausibly solve the problem of early SMBH assembly, as PBHs as large as $10^5 {\rm M}_{\odot}$ that form relatively late have not been ruled out \cite{bean02,Kawasaki:2012kn,rice17,carr18}. While these objects are constrained from forming all the dark matter (see Figure~\ref{fig:dm}), supermassive black holes are rare objects and only a small number density of black holes is necessary to explain them \cite{Clesse:2017, carr18}.
There are viable cosmological scenarios for producing PBH with masses as large as  $10^5M_\odot$~\cite{Kawasaki:2012kn,Kawasaki:2019iis}. Large PBHs could also form from dark matter self-interactions \cite{Feng:2020kxv,Feng:2021rst}. 
Constraining a primordial origin for these objects would require deep observations of high redshift quasars \cite{Sanderbeck:2021szv}. Although these objects can accrete, current CMB accretion constraints allow for the existence of enough PBHs to provide the observed high-redshift SMBH seeds \citep{Serpico:2002.10771}.

\section{Measurements of PBHs}

Numerous observations probe the population of PBHs, constraining the mass range in which PBHs are a viable dark matter candidate. The identification of PBHs remains valuable even if they provide only a minimal fraction of cosmological dark matter, so as well as closing windows for PBH dark matter, future observations will probe PBH scenarios with much smaller abundances. Many of the existing and projected constraints are reviewed in Refs.~\cite{0912.5297,Carr:2016drx,Carr:2020gox,Carr:2020xqk,Green:2020jor}, and a repository of digitized constraints is publicly available \cite{bradley_j_kavanagh_2019_3538999}. PBH observables fall broadly into a few classes: gravitational microlensing, gravitational waves, electromagnetic signatures from evaporation and accretion and dynamical effects. 
Of these, microlensing offers the most robust constraints for high mass objects, gravitational waves offer the most discovery potential and electromagnetic signatures offer unique abilities to constrain the lowest mass range.




\subsection{Gravitational Microlensing}

Gravitational microlensing, the achromatic brightening and dimming of background stars due to the transit of a massive compact foreground object, can be used to directly detect and measure the properties of PBHs.
The first \emph{direct} constraints on the compact nature of dark matter were set in the 1990s by the MACHO \citep{1992ASPC...34..193A}, OGLE \citep{1992AcA....42..253U}, and EROS \citep{1993Msngr..72...20A} collaborations. Current microlensing constraints set limits on the black hole abundance at the level of $10\%$ for black holes in the mass range $0.01 - 10 \Msun$ (see Figure~\ref{fig:dm}), and can be extended to $\sim 100 \Msun$ by combining different microlensing surveys \cite{2022arXiv220201903L}. 

Microlensing is highly complementary to GW probes since it can detect isolated or free-floating black holes \cite{Sahu:2022}.
Gravitational microlensing results in two potentially observable features: (1) photometric microlensing, a temporary amplification of the brightness of a background source, which is achromatic in the case of unblended sources, and (2) astrometric microlensing, an apparent shift in the centroid position of the source.
The characteristic photometric signal of a simple point-source, point-lens (PSPL) model as observed from the center of the solar system is symmetric, achromatic, and has both a timescale and maximum amplification that depend on the mass of the lens.
This simple PSPL model is complicated by astrophysical factors including the velocity distribution of sources and lenses, extinction due to Galactic dust, blending in dense stellar fields, stellar and planetary companions, and the shift in perspective resulting from viewing a microlensing event while the Earth revolves around the Sun.
Fortunately, these complications can be addressed and disentangled to arrive at the mass of the gravitational lens and a detection of dark matter via microlensing \citep{1405.3134,1509.04899}.
One particularly powerful feature for long-duration microlensing events results from the change in the geometric configuration of the source-lens-observer system as the Earth orbits the Sun. The change in viewing angle and distance results in a parallax effect that imposes a 1-year periodicity on top of an otherwise symmetric microlensing light curve.

\begin{figure}
\begin{center}
\includegraphics[width=0.9\textwidth]{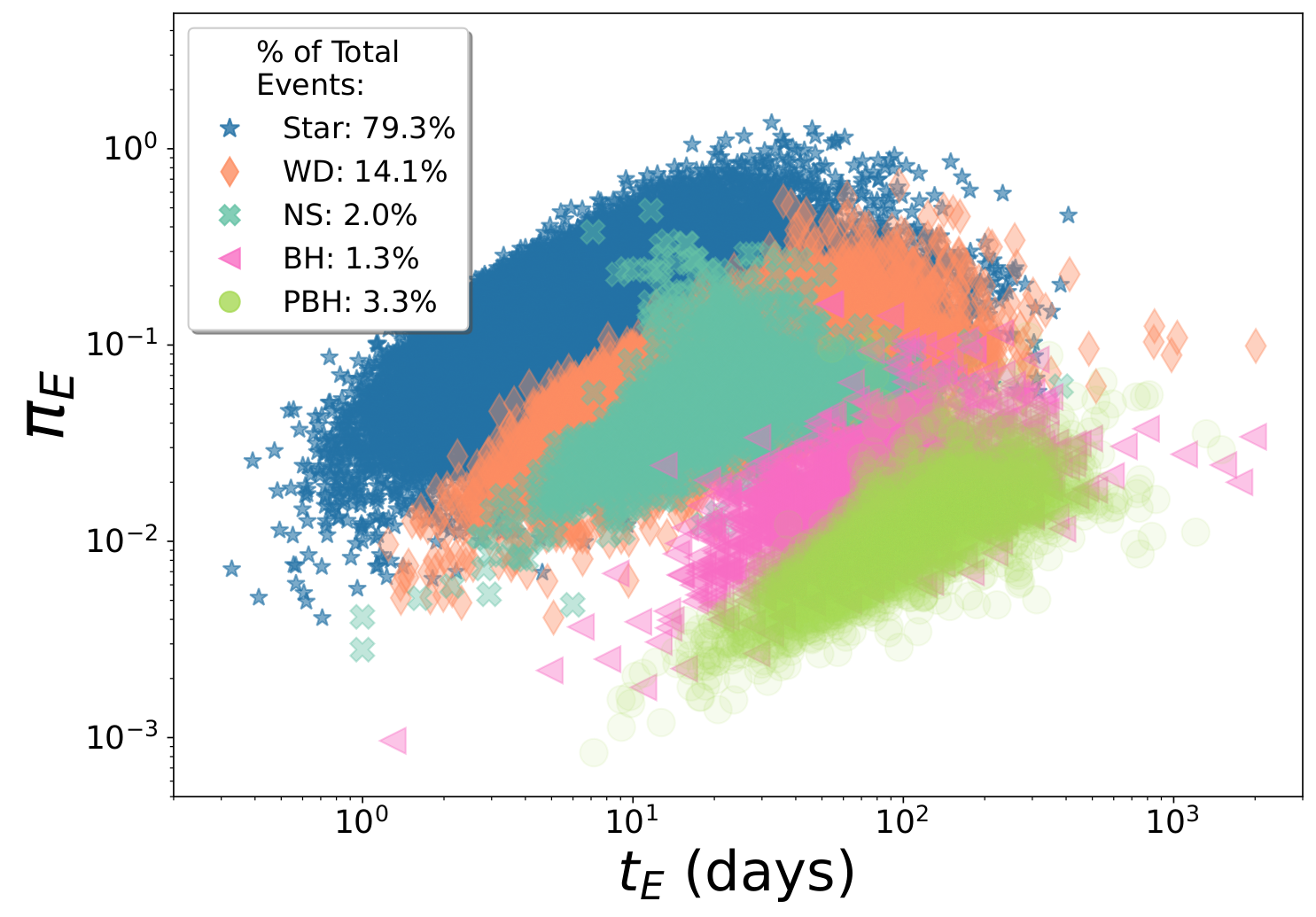}
\end{center}
\caption{A promising new means of detecting black holes with masses $\gtrsim 1\,M_\odot$ using gravitational microlensing has been identified by \cite{Lam2020}. By measuring both the duration ($t_E$) and parallax ($\pi_E$) of the microlensing event it is possible to distinguish black hole lenses from other astrophysical sources. The larger the mass of the lens the more it will move to the lower right of the parameter space and be distinguishable from other astrophysical events. Even if there is overlap with the mass of astrophysical black holes (pink triangles), we can statistically constrain the PBH population (green circles) based on relative abundance. The PBH population in this figure is composed of $35 M_\odot$ black holes distributed in an NFW Milky Way halo; approximately consistent with the log-normal component of the GW observed mass distribution \cite{LIGOScientific:2021psn}. Figure from Pruett et al.\,in prep.}
\label{fig:microlensingBH}
\end{figure}

Ref.~\cite{Lam2020} identified a promising new means of detecting black holes with masses $\gtrsim 1\,M_\odot$.
They found that black holes $\gtrsim 1\,M_\odot$ preferentially lie within a unique region of microlensing phase space, see Figure~\ref{fig:microlensingBH}, and that by measuring both the duration ($t_E$, which scales like the square root of the lens mass $t_E \sim M^{1/2}$) and parallax ($\pi_E$) of the microlensing event it is possible to distinguish black hole lenses from other astrophysical sources.
By measuring the relative abundance of microlensing events throughout this phase space and comparing them with microlensing simulations that include PBH dark matter (e.g., see Figure \ref{fig:microlensingBH} from Pruett et al. in prep.) it is possible to improve and extend microlensing dark matter constraints to masses greater than $10 M_\odot$.
In order to accurately constrain the abundance of the PBHs it is still necessary to run many such microlensing simulations to properly explore possible PBH mass distributions as well as marginalize over nuisance Milky Way parameters. LSST will provide exceptional constraints from microlensing (see Section~\ref{sec:lsst}). 

Microlensing observations using Subaru HSC have produced the strongest limits on PBH in the sublunar mass region~\cite{1701.02151,Montero-Camacho:2019jte,Smyth:2019whb,Croon:2020ouk}, as well as a candidate event consistent with a PBH of mass $\sim 10^{-7} M_\odot$~\cite{1701.02151}.  A reanalysis of OGLE data also produced several candidate events with masses $\sim 10^{-5} M_\odot$~\cite{Niikura:2019kqi}. Black holes around $1 M_\odot$ (ie, in the mass gap region) have been detected by OGLE-GAIA~\cite{Wyrzykowski:2019jyg} and may have a primordial origin. Microlensing constraints from HSC can be used to constrain potential stochastic GW background signals such as those reported by NANOGrav~\cite{Sugiyama:2020roc, Ashoorioon:2022raz}. The main systematic uncertainty is the distribution of dark matter in the Milky Way halo~\cite{Calcino:2018}, which also affects other dark matter experiments. Microlensing constraints are insensitive to small-scale PBH clustering \cite{Petac:2022rio}, but are strongly affected by large-scale PBH clustering~\cite{Carr:2019kxo}.


Another unique probe of PBHs in the $\mathcal{O}(10 -100 M_\odot)$ range is strong lensing of fast radio bursts~\cite{Munoz:2016tmg, Laha:2018zav}. Recent analysis of the first CHIME catalog indicates that with a thousand FRBs detected, a robust bound can be placed on a fraction $f=1$ of dark matter in PBHs~\cite{Zhou:2021ndx,Krochek:2021opq}. With thousands of FRBs expected to be detected over the next few years, this bound will steadily improve. Strong lensing of Gamma-Ray Bursts has also been considered as a PBH dark matter probe~\cite{Ji:2018rvg}, but  will require future instruments with improved sensitivity. 

As a purely gravitational probe, lensing is sensitive to primordial black holes even outside of galactic microlensing surveys. Indeed, the strongest current constraints on primordial black hole dark matter at $30 M_\odot$ come from lensing of supernovae at cosmological distances \cite{Zumalacarregui:2018,Garcia-Bellido:2017imq}. Other suggestions include disruption of the extreme lensing produced near strong lens caustics \cite{Venumadhav:2017pps} or lensing of quasar light curves \cite{Mediavilla:2017bok}. Future lensing catalogues obtained with high resolution imaging (e.g.~with the James Webb Space Telescope, the Roman Space Telescope or an Extremely Large Telescope) could likely be re-used for primordial black hole constraints (see Section~\ref{sec:elt}). 

\subsection{Gravitational Wave Mergers from PBHs}

Gravitational wave (GW) signals offer the highest discovery potential for PBHs. The detection of a GW merger (or GW related stochastic gravitational wave background) in a regime where non-primordial black hole formation channels are not present would be an unambiguous smoking gun signal of PBHs. For example, the detection of black hole mergers before the formation of the first stars at $z > 20$, or where one of the primaries have masses in the sub-solar range where astrophysical BHs are unlikely to form.

There are also statistical smoking guns available with a large enough sample of black hole mergers. These could include the detection of a substantial stochastic gravitational wave background from unresolved merging black holes at high redshifts ($z > 10$) \cite{Mandic:2016lcn}. Alternatively, with enough sources one could cross-correlate galaxies and black hole mergers to measure the black hole clustering signal \cite{Raccanelli:2016cud}. Primordial black hole mergers are hosted mostly by halos with masses less than the cutoff for star formation, and so cluster substantially less than electromagnetically visible stars.

Should two primordial black holes pass close enough to each other to become bound, they will be visible through their gravitational wave emission \cite{Nakamura:1997}. 
The expected PBH merger rate is theoretically hotly debated \cite[e.g.~][]{Bird:2016, Sasaki:2016, AliHaimoud:2017,Vaskonen:2019jpv,Trashorras:2020mwn, bib:Jedamzik2006.11172,Sasaki:2021iuc}. PBH binary mergers can receive significant contributions from the late Universe~\cite{Bird:2016} as well as the early Universe before matter-radiation equality~\cite{Sasaki:2016}. Large merger rates were expected from binaries formed in the early Universe \citep{AliHaimoud:2017,Sasaki:2018}. However, these merger rates are substantially reduced when three-body effects are included \cite{bib:Raidal1812.01930}, as these cause cluster formation and evaporation at high redshift \cite{Trashorras:2020mwn,bib:Jedamzik2006.11172}.  
Interpretation of GW observations crucially depends on expected merger rates and so significant theoretical work is necessary to ensure robustness of future constraints (see Section~\ref{sec:theory}). 

A priori, any GW merger event could involve a PBH as one of the progenitors \cite[e.g.~][]{Takhistov:2017bpt,Tsai:2020hpi},\footnote{Ref.~\cite{Sasaki:2021iuc} showed that PBH-neutron star mergers are subdominant to astrophysical merger rates.}
However, astrophysical black holes are known to exist and are expected to produce mergers. Black hole no-hair theorems imply that individual mergers cannot be unambiguously attributed to primordial or astrophysical sources from their gravitational wave signatures alone. Pending detection of high redshift or sub-solar mass events, distinguishing between astrophysical and primordial black holes requires a statistical study of the black hole merger population \cite{Hutsi:2020sol}. These kinds of statistical signals are low-cost; the raw data is already being gathered by the LIGO/VIRGO collaboration (and ultimately by LISA). Dedicated analysis efforts are however required to turn the data into population constraints.

Existing data has already begun to produce tantalising hints which, while by no means conclusive, are possibly signatures of PBH mergers in the current GW merger catalogue. These include the existence of a peak in the GW mass function at around $30 M_\odot$, which has become increasingly significant in recent LIGO-VIRGO data releases \cite{LIGOScientific:2021psn}. Indeed the best fit model to the LIGO-VIRGO mass function remains the combination of a power law and a Gaussian, which could be produced as a consequence of a sub-population of primordial black holes \cite{Kovetz:2017}. Another hint may come from the systematically low spins of the components of the observed BBH mergers \cite{Garcia-Bellido:2020pwq}, which are consistent with a population of primordial black holes.


\subsection{Electromagnetic signatures}
Generically PBHs are non-interacting and dark. However, there are two exceptions which can directly produce electromagnetic signatures. First, light PBHs ($\lesssim\SI{e-15}{M_\odot}$) produce significant Hawking radiation, and second, if massive PBHs ($\gtrsim\SI{10}{M_\odot}$) are able to acquire a reservoir of baryons, they can sometimes become luminous due to accretion.

\subsubsection{Evaporation bounds}

Black holes evaporate and lose mass through emission of Hawking radiation.
Since the BH temperature $T_{\rm PBH} \sim M_{\rm pl}/M$ is inversely proportional to its mass, and the evaporation timescale scales as $t_{\rm evap} \sim t_{\rm pl} (M/M_{\rm pl})^3$, lighter PBHs evaporate more efficiently. In particular, PBHs with mass $M\lesssim M_{\rm pl} (t_{\rm pl} H_0)^{-1/3} \sim 10^{14}$ g would have entirely evaporated by the present time.
They therefore cannot exist or make up any of the DM today~\cite{carr18}. However, since their evaporation takes place early, it would not be detectable in gamma rays and so their non-existence today does not imply limits on the power spectrum.

Larger mass PBHs still emit Hawking radiation. While not enough to evaporate the PBH, for some mass ranges this is sufficient to be detectable. Since evaporation emission produces a wide spectrum of particles, a variety of signatures can be associated with such PBHs of $M \sim 10^{15}$~g, including $\gamma$-rays \cite{Laha:2019ssq,Laha:2020ivk}, neutrinos or antiprotons. Radiation may also be detectable through its secondary heating effect~\cite{Kim:2020ngi,Laha:2020vhg}.

Evaporation around the epoch of recombination can modify the primordial ionization history, and as a consequence leave an imprint on and be constrained by CMB anisotropy power spectra \cite{Poulin_2017, Clark_2017, Poulter_2019}. Injection of energy through evaporation can modify the CMB blackbody or 21 cm spectrum, introducing a $\mu$ distortion or a $y$ distortion depending on the epoch during which evaporation takes place \cite{Chluba_2012,Nakama:2017xvq,Cang:2020aoo,Cang:2021owu, Saha:2021pqf}. Spectral-distortion bounds are usually much weaker than CMB anisotropy limits, however, except for PBHs evaporating well before recombination (as is the case for decaying DM \cite{Poulin_2017}). Energy injection from PBH evaporation can also significantly modify the reionization history of the universe, with associated \SI{21}{\centi\meter} signatures \cite{Villanueva-Domingo:2021cgh}. Constraints on $\Delta N_{\rm eff}$ from CMB-S4, together with measurements of a stochastic gravitational wave background, could constrain the existence of radiation decay byproducts and induced gravitational waves from primordial black holes that have decayed early on~\cite{Domenech:2021wkk}.



\subsubsection{Accretion bounds}

PBHs accrete efficiently during radiation domination and before recombination. Accreting PBHs may thus affect the cosmological ionization history, and hence CMB anisotropies \cite{Miller_2000, Ricotti:2008, AliHaimoud:2017cmb, Poulin_2017b, Serpico:2002.10771}. In principle, the energy injected by accreted PBHs may also lead to CMB spectral distortions, but these are systematically less constraining than CMB anisotropies \cite{AliHaimoud:2017cmb}. The physics of PBH accretion is complex and remains poorly understood, and different models give dramatically different accretion luminosities. Assuming a low radiative efficiency appropriate for a quasi-spherical accretion flow, one finds that PBHs with masses $M \gtrsim 100 M_{\odot}$ are conservatively ruled out as all of the DM by current CMB-anisotropy power spectrum measurements \cite{AliHaimoud:2017cmb}. If the accretion flow around PBHs is disk-like and thus significantly more efficient, CMB anisotropies could probe PBHs with masses as low as $\sim 0.1 M_{\odot}$ \cite{Poulin_2017b}. The modulation of PBH accretion by the supersonic relative velocities of baryons and DM induces large-scale fluctuations in their accretion rate, thus luminosity, which in turn implies inhomogeneous perturbations to the ionization history \cite{Jensen_2021}. This ought to lead to unique non-Gaussian signatures in CMB anisotropies \cite{Dvorkin_2013}.

Massive PBHs can acquire accretion discs by interacting with the interstellar medium even at low redshift. The emission from these accretion discs creates emission spectra across the electromagnetic spectrum, depending on the gas density and PBH mass~\cite{Lu:2019,Takhistov:2021aqx}. X-ray observations limit the number of such sources, leading to an observational bound on PBHs with masses greater than a few $M_\odot$~\cite{Inoue:2017csr}.  
Emission from accretion disks can efficiently deposit energy into the surrounding gas. 
Spinning PBHs passing through dense galactic regions may form and sustain powerful relativistic jet outflows. Outflowing jets and winds \cite{Yuan:2015}, can efficiently heat the surrounding gas, potentially leading to stronger constraints~\cite{Lu:2019,Takhistov:2021aqx,Takhistov:2021upb}.  Gas heating by accreting PBHs provides independent constraints in similar mass-ranges as the CMB.
Observing accretion at high redshift may also provide a route to detect massive PBHs serving as SMBH seeds, and could be constrained by searching for excesses in the cosmic IR and X-ray backgrounds \cite{Cappelluti:2021usg}.




Current constraints are limited by our understanding of black hole accretion. As a component of dark matter, the PBH has a high velocity relative to the ISM gas. Thus to realise the above probes, additional theory and simulation efforts are required to better refine accretion and radiation models and so derive more accurate predictions and bounds. 

\subsection{Neutron stars as PBH laboratories}
\label{sec:neutronstar}

If dark matter is made up of PBHs with masses $(10^{-16}-10^{-10}) M_\odot$, a PBH can be captured by a neutron star, transmuting (\cite{Takhistov:2017bpt}) the host star into a $\sim 1-2$ solar mass black hole~\cite{Capela:2013yf,Kouvaris:2013kra,Takhistov:2017bpt,Fuller:2017uyd,Takhistov:2017nmt,Takhistov:2020vxs,Tsai:2020hpi}. This process results in a population of black holes that can be distinguished from astrophysical black holes~\cite{Takhistov:2020vxs,Dasgupta:2020mqg}. Neutron rich material ejected from a rapidly spinning neutron star during the last milliseconds of its demise can contribute to r-process nucleosynthesis~\cite{Fuller:2017uyd} and can be observed as a kilonova by optical telescopes~\cite{Fuller:2017uyd,Bramante:2017ulk}. Instead of a merger event, the asteroid mass black hole settles inside the neutron star and gradually accretes its material. Thus the event is not expected to be accompanied by significant gravitational wave emission. Future surveys such as LSST may be able to detect kilonovae not triggered by the gravitational waves signals~\cite{DES:2017dgt}. A kilonova lacking the gravitational waves counterpart would be smoking-gun evidence of PBHs~\cite{Fuller:2017uyd,Bramante:2017ulk}.

\section{Facilities Enabling Measurements of PBHs}


There are a variety of facilities which potentially enable measurements of PBHs. In most cases the PBH science is a secondary target for well-motivated future facilities. The action items are thus data analysis and theoretical work. Theoretical work is discussed in Section~\ref{sec:theory}. Here we focus on the potential constraints available from each facility, other applications of which are discussed further in the CF03 facilities paper.

\subsection{Rubin Observatory LSST}
\label{sec:lsst}

The upcoming Rubin Observatory LSST provides an exciting opportunity to directly measure the mass function of compact object through microlensing.
Existing microlensing surveys lose sensitivity at $M \gtrsim 10 \Msun$ due to the ${\sim}5$-year duration of these surveys (although see \cite{2022arXiv220201903L} which combines two surveys to achieve sensitivity to $M \sim 100 \Msun$). While the nominal LSST has a similar duration (extensions to LSST may be considered \cite{Snowmass2021:RubinII}), it can surpass this limitation by directly detecting events using the parallactic component of the lensing signal. With this technique, LSST will observe billions of stellar sources in multiple filters over several years to enable the detection of thousands of microlensing events across a wide range of timescales and consequently a wide range of masses. 

LSST will directly detect compact halo objects through gravitational microlensing observations. It will be sensitive to both low-mass objects through short ($\sim30\,$s) events and high-mass objects through long-duration ($\sim$ years) events.
If scheduled optimally, the wide field-of-view, high cadence, and precise photometry of LSST will provide sensitivity to microlensing event rates corresponding to $\roughly 0.03\%$ of the dark matter density in compact objects with masses $>0.1\Msun$ (see Figure~\ref{fig:dm}).
Ref~\cite{drlica-wagner_2019_lsst_dark_matter} projected that LSST will be sensitive to PBH making up a fraction of $10^{-4}$ of the dark matter over a wide range of PBH masses (see also \citep{Winch:2020}).
Dedicated mini-surveys of high stellar density fields (similar to those performed with HSC \citep{1701.02151}) will yield sensitivity to much lower mass PBHs. Figure~\ref{fig:dm} shows the forecast constraints from Rubin LSST, although at high masses, $M > 10^3 M_\odot$, these constraints may be weakened by source blending.

As a time-domain survey, LSST will also be able to detect kilonovae lacking gravitational waves signals~\cite{DES:2017dgt}, smoking gun evidence for the existence of low-mass (and thus primordial) black holes, as discussed in Section \ref{sec:neutronstar}.

If compact objects make up a significant fraction of dark matter, LSST will provide insight into the primordial perturbations and early universe equation of state, in the case of PBHs, or provide evidence that dark matter particle physics is complex enough to allow significant cooling channels.
Dark matter with sufficient self-interactions to cool will also affect halo profiles, and LSST will be well-placed to distinguish different models for the formation of novel compact objects \citep{Winch:2020}.


\subsection{Roman Space Telescope}

As a high resolution space-based imaging system, the Roman Space Telescope has the potential to detect or constrain PBHs through various types of lensing. A hypothetical photometric microlensing survey of the Galactic bulge could detect PBHs comprising  $5-10$\% of the dark matter \cite{Lu:2019}.
Roman will also allow an M31 microlensing survey. As Figure~\ref{fig:dm} shows, such a survey provides the strongest constraining power for low-mass compact objects. Rubin is in the southern hemisphere and thus is unable to observe M31.

Furthermore, the high angular resolution of Roman enables the detection of astrometric microlensing. 
Astrometric microlensing relies on the fact that the two images generated during a compact object lensing event will be of differing brightness and the brightness ratio of these two images will vary throughout the duration of the lensing event. 
The two images will be most similar in brightness when the projected lens-source separation is at its minimum. 
The planned Roman exoplanet (astrometric) microlensing survey will be sensitive to PBHs at the level of about $40\%$ of the dark matter \cite{Pardo:2021uzy}. 
An astrometric microlensing survey conducted by Roman would complement LSST by breaking degeneracies between lensing mass and geometry, allowing for precise measurements of individual black hole masses, thereby measuring the black hole mass spectrum in the Milky Way halo \cite{Yee_2015}.
Finally, Roman will detect a large number of strong lenses, greatly improving current constraints from the disruption of strong lens caustics by PBHs  \cite{Venumadhav:2017pps}.

\subsection{Extremely Large Telescopes}
\label{sec:elt}

The $30$-m class extremely large telescopes (ELTs) will also be sensitive to microlensing events. 
Similar to Roman, the greatest utility from ELTs is likely to come from using astrometric microlensing to break lens degeneracies in microlensing events detected by LSST \cite{2016ApJ...830...41L,2022arXiv220201903L,2019BAAS...51c.365L}.
ELTs will be able to perform follow up observations of compact object candidates with high-precision astrometry and high spatial resolution
Future ELTs, equipped with adaptive optics (AO), will
have an astrometric precision at least an order of magnitude better than the current AO system on Keck and spatial resolutions of $<$20 mas. 
Furthermore, simultaneous measurements with a ground-based ELT and Roman in orbit at L2 will enable instantaneous space parallax measurements, which can constrains the distance ratio between the lens and source.

\subsection{LIGO}

Gravitational wave detectors such as LIGO have a high degree of discovery potential for PBHs. LIGO/VIRGO data analysis continues to search for unexpected signals. In particular, should LIGO detect black holes in a mass range unlikely to be the result of stellar evolution, this would constitute evidence for PBHs. Current data already place meaningful constraints on the merger rate of $220−24200$
${\rm Gpc}^{−3}$ ${\rm yr}^{−1}$,
depending on the chirp mass of the binary \cite{LIGOScientific:2021job}. Turning this merger rate into a PBH fraction constraint is model-dependent, and so these observations motivate the theoretical work described in Section~\ref{sec:theory}.

There is also the possibility that the binaries already detected by LIGO/VIRGO may have a primordial origin, as discussed extensively in the literature \cite{Bird:2016, Sasaki:2016, Clesse:2017}. Perhaps the most tantalising hint for this possibility comes from the peaks observed in the black hole binary mass function \cite{LIGOScientific:2021psn}, which are suggestive of multiple populations of black holes. Indeed, a recent Bayesian analysis selecting between black hole formation models showed a mild preference for a PBH component \cite{Franciolini:2021tla}. Further investigations of this possibility are clearly warranted \cite{Franciolini:2021xbq}. They require data analysis, continued theory work and the continued development of the LIGO detectors.

\subsection{Cosmic Explorer}

The next generation of gravitational wave experiments, of which Cosmic Explorer (CE) is the US component \cite{Evans:2021gyd} and Einstein Telescope (ET) the European \cite{2020JCAP...03..050M}, would increase sensitivity in the LIGO/VIRGO frequency band ($\sim 10 - 1000$ Hz) by an order of magnitude. With this improved sensitivity would come transformative ability to probe PBHs. The redshift range of the next-generation GW network would reach to $z\sim 30$. The detection of BH mergers at this redshift, before the formation of the first stars, would be clear evidence for a primordial origin.

A second avenue for discovery is the detection of a high redshift stochastic gravitational wave background from faint, unresolved, binary black hole sources \cite{2016PhRvL.117t1102M}. Ref~\cite{Ng:2021sqn} showed that redshift uncertainties mean that CE would most likely be able to securely place a handful of mergers at $z \geq 30$. The stochastic signal would provide invaluable evidence confirming any potential primordial events.

Finally, improved statistics from CE would allow detailed analysis of the present-day binary black hole mass function, allowing features in the population to be securely detected. The redshift evolution of the merger rate could be measured to $z\sim 2$ \cite{2020JCAP...03..050M}; should this evolution track the star formation rate it would be evidence for a stellar origin of the merging black holes. Should it evolve differently, there would be evidence for a primordial component.

Action items for these next-generation networks are first, to design and build the detectors over the next decade and second, as with current LIGO, continued theory work and data analysis.

\subsection{Pulsar Timing with ngVLA}

At low frequencies ($10^{-9} - 10^{-7}$ Hz),  stochastic GW backgrounds may be constrained using pulsar timing arrays. The next-generation radio interferometer ngVLA will improve GW bounds by an order of magnitude over the current constraints from the NANOGrav 12.5-year dataset \cite{2018ASPC..517..751C}. The NANOGrav collaboration has detected a stochastic signal \cite{NANOGrav:2020bcs}. It is not yet a secure gravitational signal as quadrupolar spatial correlations have not been detected. However, should it be confirmed, one interpretation is GWs induced by the formation of sub-solar mass PBHs (the mechanism discussed in Section~\ref{sec:subhorizon}) \cite{Kohri:2020qqd,PhysRevLett.126.041303,PhysRevLett.126.051303}. The order of magnitude improvement in sensitivity from ngVLA would allow confirming or ruling out the gravitational wave nature of the signal. It would also become possible to measure the slope of the background and thus distinguish a PBH explanation from other potential high energy (such as cosmic string) or astrophysical sources.

\subsection{Fast Radio Bursts observatories}

Radio experiments, such as CHIME and HIRAX, will obtain a large number of FRBs, and so allow statistical studies on their lensing, detecting possible PBH dark matter lenses. The characteristic time-delay signature of strong lensing would allow us to separate a detection from systematics, and to moreover characterize the PBH masses if lensing was detected~\cite{Munoz:2016tmg}. The upcoming Canadian Hydrogen Observatory and Radio transient Detector (CHORD~\cite{Vanderlinde:2019tjt}), the successor to CHIME, would provide a unique opportunity to detect primordial black holes through their lensing signature in the time domain, given their expectation to detect on order of tens of thousands of FRBs at large cosmological distances.

\subsection{Laser Interferometer Space Antenna} 

The space-based LISA mission will be sensitive to gravitational waves in a frequency band from $\sim 10^{-5} - 0.1$ Hz. As with any gravitational wave experiment, there is discovery potential by detecting PBHs which merge in this frequency band. A unique feature of LISA is its ability to detect extreme mass ratio inspirals, when a low-mass black hole merges with a much more massive object. In some models of PBH formation with a wide mass function, the expected rate of these mergers can be larger than $10^3$ yr$^{-1}$ Gpc$^{-3}$ \cite{LISAwhitepaper}.

LISA's strain sensitivity will allow it to constrain the stochastic GW background sourced by the subhorizon formation of PBHs \cite{Bartolo:2016ami,PhysRevLett.122.201101, Barausse:2020rsu,Braglia:2021wwa}. LISA would also be able to detect the stochastic signal from NANOGrav, if it has a PBH origin, as it is expected to be almost flat from $10^{-8}$ Hz to $10^{-2}$ Hz \cite{PhysRevLett.126.051303}.

The astrophysical range of LISA will allow it to probe the existence of intermediate mass binary black hole mergers at redshift $z > 20$ with a SNR larger than five, for equal-mass mergers and progenitor masses between $10^{3} M_\odot$ and $10^{6} M_\odot$ \cite{LISAwhitepaper}. This would constitute a probe of the formation scenario for SMBHs described in Section~\ref{sec:smbh}, where a collection of PBHs merge at high redshift, before the formation of astrophysical black holes, and seed the first quasars.

\subsection{Mid-band GW Detectors: AEDGE}

There is a frequency gap from around $0.01 - 10$ Hz between the ground-based (LIGO/Cosmic Explorer) and space-based (LISA) gravitational wave experiments. This gap can be filled with a midband experiment, either using LISA-like technology but shorter arms \cite[e.g.~DECIGO][]{Sato:2017dkf} or an atomic interferometer \cite{AEDGE:2019nxb}. The unique capability of a midband experiment is the detection of intermediate mass ratio mergers, such as those expected during the early stages of supermassive black hole assembly. Detection of these mergers at sufficiently high redshift that astrophysical black holes have not yet formed would constitute evidence for PBHs. Furthermore, closing this frequency gap would strengthen the ability of all GW detectors to measure stochastic gravitational wave backgrounds \cite{Barish:2020vmy}. Continuous frequency coverage allows for better component separation, making it easier to distinguish astrophysical and high energy signals.

\subsection{MeV Gamma-Ray Telescopes}

The MeV gamma-ray band remains under-explored compared to the neighboring GeV and X-ray bands, and is particularly important for probing asteroid-mass PBHs.
That is because PBHs evaporating at present with lifetimes near the age of the universe have masses near $5\times 10^{-19}\, \Msun$, lying at the lower end of the asteroid-mass gap, and thus generically emit gamma rays with energies of $\mathcal{O}(10\, \mathrm{keV}) - \mathcal{O}(1\, \mathrm{MeV})$.
The radiation from such PBHs could contribute to the galactic and extragalactic gamma-ray spectrum, and would have a distinct spectrum from astrophysical backgrounds.

COMPTEL observations of the Milky Way dating back two decades give some of the strongest constraints on the abundance of PBHs in the $2.5\times 10^{-18} - 5\times 10^{-17}\, \Msun$ mass range~\cite{Coogan:2020tuf}.
Likewise, INTEGRAL gamma-ray data of the masked Galactic profile has set the tightest bound on the mass of PBHs that are allowed to compose the DM~\cite{Laha:2020ivk}.
Looking at the future, several telescopes have been proposed to fill the MeV gamma-ray sensitivity gap, including AdEPT~\cite{adept}, AMEGO~\cite{amego, Fleischhack:2021mhc}, All-Sky ASTROGAM~\cite{as_astrogam}, GECCO~\cite{gecco}, GRAMS~\cite{grams,grams_loi}, MAST~\cite{mast} and PANGU~\cite{pangu,pangu_aeff}. 
Observations of galactic and extragalactic targets with several of these telescopes have the potential to extend constraints on PBH dark matter from $\sim 5\times 10^{-17}\, \Msun$ to $1.5\times 10^{-15}\, \Msun$, or to discover them~\cite{Coogan:2020tuf,Coogan:2021rez,Ray:2021mxu, Ghosh:2021gfa}.
As PBHs become more massive their emission dims and shifts to lower energies.
Therefore, telescopes with a low energy threshold are particularly well-suited to extending the PBH dark matter discovery reach into the asteroid-mass gap, where few other probes are available.

\subsection{GeV and TeV Gamma-Ray Observatories}
PBHs with masses of $\sim 5 \times10^{-19}M_{\odot}$ would be currently evaporating and are expected to produce TeV gamma rays. These would be detected at the end of the PBH's life as a burst of gamma rays. We expect these bursts to occur isotropically across the sky. Therefore, a wide field of view survey observatory is ideal for these searches. For example, the {\it Fermi}-LAT and HAWC can observe 15\% of the sky at any given moment. These instruments also search the entire sky (or close to the entire sky) with a high duty cycle of at least daily. 
Recent searches have been performed by the {\it Fermi}-LAT \cite{FermiPBH1,FermiPBH2} and HAWC \cite{HAWCPBH} observatories. The best limits come from HAWC with a burst rate density upper limit of ${\sim} 3\times 10^3 pc^{-3} yr^{-1}$. Future wide field of view survey observatories with increased sensitivity are being proposed. For example, SWGO is expected to set an upper limit on the burst rate density of  ${\sim} 50 pc^{-3} yr^{-1}$, thus increasing the current sensitivity by almost two orders of magnitude after 5 years of observation\cite{SWGOPBH}.

Existing imaging atmospheric Cherenkov telescopes, such as VERITAS, MAGIC, and H.E.S.S., despite their narrow field of view ($\sim5^{\circ}$ in diameter), have excellent sensitivity for TeV gamma-ray transients and can provide competitive upper limits on the burst rate density. For example, the strongest 95\% upper limits are $527 \;\text{pc}^{-3} \;\text{yr}^{-1}$ from 4924 hours of H.E.S.S observations \cite{Tavernier2021HESSPBH}. 
With its wider field of view and better sensitivity, the Cherenkov Telescope Array is expected to improve the constraints on the burst rate density to ${\sim} 10^2 \;\text{pc}^{-3} \;\text{yr}^{-1}$ \cite{Cassanyes2015MAGIC_CTA_PBH, Doro2021IACT}. 

\section{Theoretical Work}
\label{sec:theory}

There are a number of open questions in the theory of primordial black hole formation as well as observational aspects. While some possible detectable signatures of PBHs are well explored, other aspects of their life cycle remain obscure. Especially uncertain and important questions are: 
\begin{itemize}
    \item The rate at which PBHs merge and produce gravitational wave signatures.
    \item The connection between the initial mass function of the primordial black holes and the observable mass function.
    \item The distribution of PBH spins.
    \item Observational signatures of different PBH formation mechanisms.
\end{itemize}


\subsection{The PBH merger rate}
\label{sec:gwmerger}

Lacking any other mechanism to dissipate angular momentum, PBHs merge primarily through gravitational wave emission. This process takes place when two PBHs pass by sufficiently close at sufficiently low relative angular momentum to become bound by gravitational wave emission \cite{Peters:1963}. To be observable, these mergers must happen at a redshift low enough to be detectable by LIGO/VIRGO or LISA. Several semi-analytic estimates of the merger rate exist in the literature \cite[e.g.~][]{Nakamura:1997,Bird:2016, Sasaki:2016, bib:Jedamzik2006.11172,Sasaki:2021iuc}. However, these estimates are mostly at the order of magnitude level. 

Constraints are known to depend on the small-scale density distribution of PBHs, potentially evolving as a sub-dominant component in a background of cold dark matter. This is currently highly uncertain as the relevant mass scales are $\lesssim 100 M_\odot$, whereas N-body simulations are more commonly used to probe halos potentially forming stars $M \gtrsim 10^9 M_\odot$. Note that it is not sufficient to simulate a smaller box as mass transfer effects couple small and large scales. Furthermore, N-body simulations smooth (`gravitationally soften') the density field on small scales to avoid the formation of artificial binaries. They also generally do not include models for the GW emission of close PBH binaries.

A number of works have numerically studied the enhanced clustering induced by PBHs at high redshifts ($z\gtrsim100$).  If PBHs make up the majority of the CDM, then they can form halos potentially containing thousands of PBHs \citep{Inman:2019wvr}.  Such halos grow via hierarchical clustering, but can also evaporate from encounters between PBHs \cite{bib:Jedamzik2006.11172,bib:DeLuca2009.04731} (the latter effect is not included in the simulations of \citep{Inman:2019wvr} due to gravitational softening).  The formation of these early halos could affect various PBH constraints. 
On the other hand, if PBHs are just a fraction of the CDM then it is necessary to model the rest of the dark matter as well. A natural choice is a minimal model of cold, collisionless, dark matter and its impact on PBH binaries has been studied in \citep{bib:Kavanagh1805.09034}. If the PBH is by itself, a steep CDM halo forms \citep{bib:Adamek1901.08528,Inman:2019wvr} which leads to stronger CMB constraints as the additional mass causes increased gas accretion \citep{Serpico:2020ehh}.
Future cosmological simulations could more directly treat the evolution of PBH binaries, or evolve larger volumes to lower redshifts and so assess how PBH clusters are impacted by (or themselves impact) the formation of the first galaxies. Other avenues for investigation would be simulations of the accretion of gas onto PBHs with $M>10^4M_\odot$, or the growth of PBH seeds into SMBHs via runaway mergers.


Despite these technical challenges, some pioneering studies have simulated the discrete behaviour of PBHs in a small region less than $1$ Mpc across for $z \geq 99$ \cite{Inman:2019wvr}. The particle load of this simulation was relatively small, and a concerted simulation campaign could likely increase the size of a simulation to the point where it could make predictions for GW emission. Subgrid models could also be developed to model GW emission and the effects of tight PBH clumps. Exploration of more complex formation channels (e.g. involving multi-body interactions and dynamics) is an essential next step to definitively evaluate theoretical predictions and  merger rates.

\subsection{The Predicted Spin of PBHs}

\begin{table}[ht]
    \centering
\begin{tabular}{|l|l|l|}
\hline
    {\bf Formation mechanism} & {\bf Mass range} & {\bf PBH spin}  \\
\hline
  Inflationary perturbations~\cite{Green:2020jor}
   & DM, LIGO, supermassive & small \\
\hline
Inhomogeneous baryogenesis~\cite{Dolgov:1992pu,Dolgov:2008wu,Hasegawa:2018yuy,Kawasaki:2019iis} & LIGO, supermassive & small \\ \hline
Yukawa “fifth force”~\cite{Flores:2020drq,Flores:2021tmc} &
DM, LIGO, supermassive &  small \\ \hline 
Supersymmetry, Q-balls, no long-range~\cite{Cotner:2016cvr,Cotner:2017tir,Cotner:2019ykd} & DM ($10^{-16}-10^{-6} M_\odot$) & 
large \\ \hline
Supersymmetry, long-range scalar forces~\cite{Flores:2021jas} & 
DM ($10^{-16}-10^{-6} M_\odot$) &  small \\ \hline
 Light scalar Q-balls (not SUSY)~\cite{Cotner:2017tir} &
DM, LIGO, supermassive & large \\ \hline
Oscillons from the inflaton~\cite{Cotner:2018vug} &
DM, LIGO, supermassive & large \\ \hline
 Multiverse bubbles~\cite{Garriga:2015fdk,Deng:2017uwc,Kusenko:2020pcg} & DM, LIGO, supermassive & 
small \\ \hline 
\end{tabular}    
    \caption{Masses and spins of PBHs from different scenarios~\cite{Green:2020jor,Dolgov:1992pu,Dolgov:2008wu,Garriga:2015fdk,Deng:2017uwc,Cotner:2016cvr,Cotner:2017tir,Cotner:2018vug,Hasegawa:2018yuy,Kawasaki:2019iis,Cotner:2019ykd,Flores:2020drq,Flores:2021tmc,Flores:2021jas}}
    \label{tab:mass_spin}
\end{table}

Black holes have two properties: mass and spin, both observable through properties of the gravitational waveform as they merge. The distribution of spins thus in principle encodes substantial useful information on the origin and evolution of the black holes. If PBHs form through the collapse of an over-density (the mechanism outlined in Section~\ref{sec:horizon}) they have negligible initial spin \cite{Chiba:2017}, although more complex sub-horizon formation scenarios such as those outlined in Section~\ref{sec:subhorizon} often lead to higher spins  \cite{Cotner:2019ykd,Harada:2017fjm,Flores:2021tmc}. PBH in dense clusters scatter off each other, and close encounters could significantly increase their otherwise low spin \cite{Jaraba:2021ces}.


A relatively well-measured quantity from the gravitational waveforms is the mass-weighted effective spin of the incoming binaries aligned with their relative angular momentum, $\chi_{eff}$. The distribution of effective spins in the LIGO/VIRGO black hole population is centered around zero, indicating that the spins of the progenitor black holes are not correlated with each other \cite{LIGOScientific:2021psn}. This is a generic signal of a dynamical origin for the black hole binaries and compatible with a primordial black hole origin for some of them \cite{Garcia-Bellido:2020pwq}. Further measurements of the black hole spin distribution, especially with the upcoming increase in the size of the network to five detectors, could potentially constrain differences in the spin distribution expected from different formation models (see Table~\ref{tab:mass_spin}). 

\subsection{Formation Mechanisms and Connection to Supersymmetry and the Early Universe}


It is important to map out the landscape of cosmological scenarios for PBH formation.  Some formation scenarios discovered recently predict the correct abundance of PBH dark matter by relating the PBH abundance to the baryon asymmetry of the universe~\cite{Flores:2020drq,Flores:2021jas,Wu:2021gtd}.  It was also shown that dark matter in the form of PBHs naturally appears in the models with supersymmetry~\cite{Cotner:2016cvr,Cotner:2019ykd,Flores:2021jas}, in which case the mass scale of PBHs, set by the Planck mass and the scale of supersymmetry breaking, falls inside the open window for PBH dark matter. The properties of the present-day PBH population, such as clustering, spins, and mass function, depend on their formation history \cite{Carr:2019kxo}.  It is, therefore, important to explore the range of formation scenarios and their astrophysical predictions. Conversely, by measuring the clustering and spin of primordial black holes, one can hope to use PBH as a new window on the early universe. 

\subsection{Accretion and emission from PBHs in different astrophysical environments}

The dynamics of accretion rates, emission and outflows around black holes can be notoriously challenging to model. PBHs have velocity and spatial distributions distinct from astrophysical black holes, as well as potentially very broad mass-ranges extending over many orders of magnitude. Some accretion regimes and their associated emission will only appear in the context of PBHs but not astrophysical ones (e.g. accretion onto (sub)solar-mass BHs). Hence, PBH accretion and emission regimes require reinvigorated simulation efforts beyond what already exists for regimes associated with typical astrophysical black holes and their environments. Advancement across this frontier could be especially beneficially for definitively establishing the role of PBHs in the stellar-mass range for LIGO/VIRGO GW observations, since several of the existing constraints there (e.g. CMB) sensitively rely on the description of the above processes.

\section{Conclusion}

As we have outlined in the white paper, by constraining the abundance and properties of PBHs we can place some of the tightest constraints on the early universe, extending the CMB and gamma-ray constraints to even earlier times (to the highest wavenumbers observable). These same observations of the PBH phase-space can enable us to constrain the abundance of PBHs and determine their viability as a dark matter candidate. The science impacts of these analyses are not limited to PBHs, but in many cases also apply to other exotic compact objects (e.g.~axion stars or minihalos).

Fortunately many of the facilities able to make these observations are already well-justified (and often under construction).
For example, gravitational wave observations are already being taken by the LIGO, Virgo, and KAGRA scientific collaborations, as well as pulsar timing arrays such as NANOGrav. Rubin/LSST will be capable of making the necessary microlensing observations. Telescope facilities such as the Nancy Grace Roman Space Telescope and the Extremely Large Telescopes will be complementary.  

Much of the necessary raw data will thus exist in the coming decade.
What will not exist without support are the necessary pipelines and algorithms to turn this raw data into the required scientific data products.
Similarly, without support for new theoretical studies it will not be possible to translate these observations into scientific constraints. Already, there has been confusion in the literature regarding existing dark matter and power spectrum constraints, which could be avoided with a stronger theoretical understanding.

Many of the facilities necessary to make the initial measurements are already planned. However, the next generation of facilities need support now. The necessary microlensing observations will be well facilitated by Roman and Rubin, but there are major gains to be had with future GW facilities. In rough order of scheduled start time, these are ngVLA, LISA, Cosmic Explorer, and AEDGE/DECIGO. Gamma ray observations will require MeV gamma-ray facilities, and imaging air Cherenkov telescopes.

Many of these recommendations echo those in other white papers. For example, dark matter beyond the standard model \cite{Snowmass2021:CF07CosmicProbes}, inflation \cite{Snowmass2021:Inflation}, early universe gravitational wave probes \cite{Caldwell:2022qsj},  the CF03 facilities paper \cite{Snowmass2021:CF03Facilities}, the CF03 simulations paper \cite{Snowmass2021:CF03Simulations} and the CF03 halos paper \cite{Snowmass2021:CF03Halos}. Future gravitational wave facilities are discussed in \cite{Ballmer:2022uxx}. Particular areas of agreement are the focus on gravitational waves, and the need for theory work and data analysis.

\section{Acknowledgements}
Part of this work was performed under the auspices of the U.S. Department of Energy by Lawrence Livermore National Laboratory under Contract DE-AC52-07NA27344 and was supported by the LLNL-LDRD Program under Project No. 22-ERD-037. This document was prepared as an account of work sponsored by an agency of the United States government. Neither the United States government nor Lawrence Livermore National Security, LLC, nor any of their employees makes any warranty, expressed or implied, or assumes any legal liability or responsibility for the accuracy, completeness, or usefulness of any information, apparatus, product, or process disclosed, or represents that its use would not infringe privately owned rights. Reference herein to any specific commercial product, process, or service by trade name, trademark, manufacturer, or otherwise does not necessarily constitute or imply its endorsement, recommendation, or favoring by the United States government or Lawrence Livermore National Security, LLC. The views and opinions of authors expressed herein do not necessarily state or reflect those of the United States government or Lawrence Livermore National Security, LLC, and shall not be used for advertising or product endorsement purposes.

\bibliographystyle{JHEP.bst}
\bibliography{main.bib}

\end{document}

%% file: commands.tex
\usepackage{soul} 
\usepackage{amsmath}
\usepackage{amssymb}
\usepackage{xspace}
\usepackage{xifthen}

\mathchardef\mhyphen="2D

\newcommand{\roughly}{\ensuremath{ {\sim}\,} }

\newlength{\dhatheight}

\newcommand{\unit}[1]{\ensuremath{\mathrm{\,#1}}\xspace}

\newcommand{\Msun}{\unit{M_\odot}}

\providecommand\physrep{\ref@jnl{Phys.~Rep.}}%
\providecommand\apjs{\ref@jnl{ApJS}}%
\providecommand{\jcap}{\ref@jnl{JCAP}}%

%% file: authors.tex
\author[]{Simeon Bird}
\affil[]{University of California, Riverside}
\author[]{Andrea Albert}
\affil[]{Los Alamos National Laboratory}
\author[]{Will Dawson}
\affil[]{Lawrence Livermore National Laboratory}

\author[]{Yacine Ali-Haimoud}
\affil[]{Center for Cosmology and Particle Physics, Department of Physics, New York University, New York, NY}
\author[5,6]{Adam Coogan}
\affil[5]{Département de Physique, Université de Montréal, 1375 Avenue Thérèse-Lavoie-Roux, Montréal, QC H2V 0B3, Canada}
\affil[6]{Mila -- Quebec AI Institute, 6666 St-Urbain, \#200, Montreal, QC, H2S 3H1}
\author[7,8,9]{Alex Drlica-Wagner}
\affil[7]{Fermi National Accelerator Laboratory, P. O. Box 500, Batavia, IL 60510, USA}
\affil[8]{Kavli Institute for Cosmological Physics, University of Chicago, Chicago, IL 60637, USA}
\affil[9]{Department of Astronomy and Astrophysics, University of Chicago, Chicago, IL 60637, USA}
\author[17]{Qi Feng}
\affil[17]{Department of Physics and Astronomy, Barnard College, Columbia University, NY 10027, USA}
\author[10]{Derek Inman}
\author[8]{Keisuke Inomata}
\author[11]{Ely Kovetz}
\affil[11]{Physics Department, Ben-Gurion University of the Negev, Beersheba, Israel}
\author[10,12]{Alexander Kusenko}
\affil[12]{Department of Physics and Astronomy, University of California, Los Angeles, California 90095-1547, USA}
\affil[10]{ Kavli Institute for the Physics and Mathematics of the Universe (WPI), The University of Tokyo Institutes for Advanced Study, The University of Tokyo, Kashiwa, Chiba 277-8583, Japan}
\author[13,14]{Benjamin~V.~Lehmann}
\affil[13]{Department of Physics, University of California, Santa Cruz, Santa Cruz, CA 95064, USA}
\affil[14]{Santa Cruz Institute for Particle Physics, Santa Cruz, CA 95064, USA}
\author[15]{Julian B.~Mu\~{n}oz}
\affil[15]{Center for Astrophysics, Harvard \& Smithsonian, 60 Garden St, Cambridge, MA, 02138, USA}
\author[18]{Rajeev Singh}
\affil[18]{Institute  of  Nuclear  Physics  Polish  Academy  of  Sciences,  PL-31-342  Krak\'ow,  Poland}
\author[10]{Volodymyr Takhistov}
\author[16]{Yu-Dai Tsai}
\affil[16]{University of California, Irvine}
